\documentstyle[12pt]{article}

\begin{document}

\font\ninerm = cmr9

\baselineskip 14pt plus .5pt minus .5pt

\def\footnoterule{\kern-3pt \hrule width \hsize \kern2.6pt}

\hsize=6.0truein
\vsize=9.0truein
\textheight 8.5truein
\textwidth 5.5truein
\voffset=-.4in
\hoffset=-.4in

\title{Vortex Solutions of Nonrelativistic Fermion and Scalar
Field Theories Coupled to Maxwell-Chern-Simons Theories}
\author{Bom Soo Kim\thanks{e-mail : bskim@theory.yonsei.ac.kr},
Hyuk-jae Lee\thanks{e-mail : hjlee@theory.yonsei.ac.kr},
  and Jae Hyung Yee\thanks{e-mail : jhyee@phya.yonsei.ac.kr} \\
Department of Physics and Natural Science Research Institute \\
Yonsei University,
Seoul, 151-742, Korea}

\maketitle
\begin{abstract}
We have constructed nonrelativistic fermion and scalar field theories coupled to a Maxwell-Chern-Simons gauge field
which admit static multi-vortex solutions. This is achieved
by introducing a magnetic coupling term in addition to the
usual minimal coupling.

\end{abstract}

\baselineskip=16pt plus 3pt minus 3pt

\pagestyle{plain}
\pagenumbering{arabic}
\setcounter{page}{1}

\pagebreak[3]
\setcounter{equation}{0}
\newpage

\section{Introduction}
Since the introduction of the Chern-Simons
action as a new possible gauge field
theory in $(2+1)$-dimensional space-time \cite{Tem}, it has been successfully
applied to explain various $(2+1)$-dimensional
phenomena including the high $T_c$ superconductivity and the
integral and fractional quantum Hall effects.
The Chern-Simons term has also made it possible to construct various
field theoretic models which possess classical vortex solutions with various physically
interesting properties. They include the relativistic \cite{Jackiw},
nonrelativistic \cite{Pi} scalar field theories, and relativistic fermion field theory
\cite{Chokim} interacting with Abelian Chern-Simons fields, which admit static
multi-vortex solutions saturating the Bogomol'nyi bound \cite{Bog} that reduces
the second-order field equations to the first-order ones.
In $(2+1)$-dimensions, there are two possibilities for the kinetic term of gauge field, Maxwell and Chern-Simons
terms. However, the simple theories with both kinetic terms failed to have vortex solutions.
In order for such theories to have vortex solutions one need to add new fields or
extend the models to supersymmetric theories \cite{Minn}.

Recently, Kogan and Stern\cite{kogan} introduced a new coupling of the electromagnetic field to the matter field
interpreted as a generalization of the familiar Pauli magnetic moment coupling in $(2+1)$-dimensions.
The tensor structure in $(2+1)$-dimensions allows magnetic moment couplings independent of spin representation.
This coupling gives rise to consistent vortex solutions for the relativistic matter field theories interacting with
Abelian and non-Abelian gauge field with both Maxwell(Yang-Mills) and Chern-Simons kinetic terms\cite{Antil, eight}.

Such vortex solutions of $(2+1)$-dimensional field theories have attracted much attention
recently in the hope of understanding the dynamics of magnetic vortices appearing in high
temperature superconductors \cite{today} in the field theoretic approach \cite{manton}.
The phenomena of magnetic vortices in high temperature superconductors are determined by
electrons confined to move in 2-dimensional surfaces. To understand such phenomena in the field
theoretic approach, therefore, one is better to study the nonrelativistic fermion
field theory models which possess static magnetic vortex solutions. As a first step to understand
the dynamics of such realistic magnetic vortices in the field theoretic approach, we construct
nonrelativistic fermion field theory models that possess static vortex solutions in this paper.

In section II we construct nonrelativistic spinor field models where a single two-component
spinor field couples to a Maxwell-Chern-Simons gauge field, and show that they possess static
magnetic vortex solutions. In section III, we consider a nonrelativistic scalar field theory
coupled to a Maxwell-Chern-Simons field that supports the static vortex solutions.
We conclude with some discussions in the last section.

\section{Nonrelativistic Fermion Models }

There exist many fermonic field theory models that possess static vortex solutions. Recently
Duval, Horv\'athy and Palla \cite{duval} and N\'emeth \cite{Nemeth} have constructed nonrelativistic
spinor field models that support static vortex solutions, where a couple of two-component spinor
fields couples to a Chern-Simons gauge field. In this section we are interested in constructing
nonrelativistic spinor field models where a single two-component spinor field couples to
a Maxwell-Chern-Simons gauge field. For such models to support static vortex solutions one need to
introduce a magnetic coupling as explained in the last section.

For this purpose we follow the procedure of Levy-Leblond \cite{levy}, which enables one to write
the Schr\"odinger equation as a first order differential equation.
The nonrelativistic fermion fields must satisfy the Schr\"odinger equation,
\begin{equation}
{\cal S} \Psi = \Bigl{(} i \frac{\partial}{\partial t} + \frac{1}{2m} \vec{\nabla}^2 \Bigr{)} \psi
=  \Bigl{(} E - \frac{\vec{p}^2}{2m} \Bigr{)} \psi=0.
\label{Aeq}
\end{equation}
To write this second-order differential equation as a first-order one we introduce an operator $\Theta$ :
\begin{equation}
\Theta \psi \equiv \Bigl{(} AE + \vec{B}\cdot\vec{p} + C \Bigr{)} \psi = 0,
\label{lineareq}
\end{equation}
where $A, \vec{B}, C$ are matrices to be determined.

For the solution of Eq.(\ref{lineareq}) to obey the Schr\"odinger equation (\ref{Aeq}),
there must exist some operator $\Theta^{'} = A^{'}E + \vec{B}^{'} \cdot \vec{p} + C^{'}$,
such that operating $\Theta^{'}$ on Eq.(\ref{lineareq}) we recover the Schr\"odinger equation:
\begin{equation}
\Theta^{'} \Theta = 2m {\cal{S}}.
\label{thetaeq}
\end{equation}
By identifying the various monomials in $E$ and $\vec{p}$,
we obtain the following set of conditions:
\begin{equation}
\parbox{50mm}{
\begin{eqnarray}
A^{'} A = 0  \nonumber \\
A^{'} C + C^{'} A = 2m  \nonumber \\
C^{'} C = 0 \nonumber
\end{eqnarray}}
\nonumber
\parbox{50mm}{
\begin{eqnarray}
A^{'} B_i + B_{i}^{'} A = 0  \nonumber \\
B_{i}^{'} B_{j} + B_{j}^{'} B_{i} = -2 \delta_{ij} \nonumber\\
C^{'} B_{i} + B_{i}^{'} C = 0, \nonumber
\end{eqnarray}}
\label{Aconditions222}
\end{equation}
where $i, j = 1, 2$.
Defining the new matrices as
\begin{eqnarray}
B_{3} = i \Bigl{(} A + \frac{1}{2m} C \Bigr{)}&,&
B_{3}^{'} = i \Bigl{(} A^{'} + \frac{1}{2m} C^{'} \Bigr{)}, \nonumber \\
B_{4} = A - \frac{1}{2m} C &,&
B_{4}^{'} = A^{'} - \frac{1}{2m} C^{'},
\label{definedop}
\end{eqnarray}
the conditions (\ref{Aconditions222}) can be written as
\begin{equation}
B_{\alpha}^{'} B_{\beta} + B_{\beta}^{'} B_{\alpha} = -2 \delta_{\alpha\beta}, ~~(\alpha, \beta = 1 ~ to ~ 4).
\label{condition22}
\end{equation}
All the representation of such an algebra can be obtained from those of a Clifford algebra with dimension 2:
\begin{equation}
\sigma_{\mu} \sigma_{\nu} + \sigma_{\nu} \sigma_{\mu} = 2 \delta_{\mu\nu}, ~~(\mu, \nu = 1, 2, 3).
\label{sigmamatrix}
\end{equation}
One can easily show that
\begin{equation}
B_{\mu} = B \sigma_{\mu}, ~~~~ B_{\mu}^{'} = - \sigma_{\mu} B^{-1}, ~~~~
B_{4} = -i\beta,~~~~ B_{4}^{'} = -i\beta^{-1},
\label{assigneq}
\end{equation}
satisfy the condition (\ref{condition22}), where $B$ is an arbitrary nonvanishing constant and
$\beta$ is an arbitrary nonsingular matrix.

Since we are only interested in the irreducible representations, we may use the standard results for
the Pauli matrix algebra, Eq.(\ref{sigmamatrix}).
Thus we take $B = 1$, $B_{\mu} = \sigma_{\mu} (\mu =  1, 2, 3)$, and the nonsingular matrix $\beta$ to be
\begin{eqnarray}
\beta = \left( \begin{array}{cc} 1 & 0 \\ 0 & 1 \end{array} \right) ~~or~~
\beta = \left( \begin{array}{cc} -1 & 0 \\ 0 & -1 \end{array} \right).
\label{twobetachoice}
\end{eqnarray}
We now consider the first possibility of Eq.(\ref{twobetachoice}),
$\beta = \left( \begin{array}{cc} 1 & 0 \\ 0 & 1 \end{array} \right)$, so that
$B_{4} = \left( \begin{array}{cc} -i & 0 \\ 0 & -i \end{array} \right) $.
From this choice we obtain the matrices $A$ and $C$ :
\begin{eqnarray}
A = \left( \begin{array}{cc} -i & 0 \\ 0 & 0 \end{array} \right),~~~~
C = \left( \begin{array}{cc} 0 & 0 \\ 0 & 2im \end{array} \right).
\label{ACequation11}
\end{eqnarray}
By using these results the first-order Schr\"odinger equation (\ref{lineareq}) can be written as
\begin{equation}
\Bigl{(} \frac{1}{2}(1+\sigma_3 )E + i\vec{\sigma}\cdot\vec{p} - m(1-\sigma_3 ) \Bigr{)} \psi = 0.
\label{nonmomentumeq1}
\end{equation}
In coordinate space representation Eq.(\ref{nonmomentumeq1}) reads
\begin{equation}
\Bigl{(} \frac{i}{2}  (1+\sigma_3 )\partial_t + \vec{\sigma}\cdot\vec{\partial} - m(1-\sigma_3 ) \Bigr{)} \psi = 0,
\label{nonrelativisticeq111}
\end{equation}
where $\partial_t = \frac{\partial}{\partial t}$ and $\partial_i = \frac{\partial}{\partial x_i}$.

The other choice of nonsingular matrix $\beta$ of Eq.(\ref{twobetachoice}) enables one to construct another
first-order differential
equation which is independent of Eq.(\ref{nonrelativisticeq111}). We will discuss the difference between
the two choices later.

We can easily construct a Lagrangian which leads to the equation of motion (\ref{nonrelativisticeq111}).
Introducing a coupling to the Maxwell-Chern-Simons field and a four-fermion self interaction term,
the Lagrangian for the nonrelativistic spinor field can be written as
\begin{eqnarray}
{\cal L}=-\frac{1}{4}F_{\mu\nu}F^{\mu\nu}
+\frac{\kappa}{4}{\epsilon}^{\mu\nu\rho}A_{\mu}F_{\nu \rho}
&+&{\psi}^{\dagger}\Bigl{(}\frac{i}{2}(1+\sigma_{3}){\cal{D}}_{t}
+\vec{\sigma}\cdot\vec{\cal{D}}-m(1-\sigma_3)\Bigr{)}\psi   \nonumber \\
&+&g{\Bigl{(}{\psi}^{\dagger}\frac{1}{2}(1+\sigma_3)\psi\Bigr{)} }^2 ,
\label{lagrangian1}
\end{eqnarray}
where $\psi = \bigl{(}
\begin{array}{ll}
{\psi}_{1} \cr {\psi}_{2}   \nonumber
\end{array} \bigr{)} $
is a two-component Pauli-spinor field, ${\sigma}_{i}$'s denote the Pauli matrices, and
$\kappa$ and $g$ stand for the Chern-Simons coupling constant and non-linear self-interaction
constant, respectively. In Eq.(\ref{lagrangian1})  ${\cal{D}}_{\mu}$ represent the covariant derivatives,
\begin{eqnarray}
{\cal{D}}_{t} &=& {\partial_{0}} + ieA_{0} +i \frac{l}{2} \epsilon^{ij}F_{ij},
\label{cov-deri0}  \\
{\cal{D}}_{i} &=& {\partial_{i}} + ieA_{i},
\label{cov-deri1}
\end{eqnarray}
where the last term of Eq.(\ref{cov-deri0}) represents the anomalous magnetic coupling,
$ i, j = 1, 2$ and we have used the Minkowski metric notation with signature $(+,-,-)$.

With both kinetic terms, Maxwell and Chern-Simons, the field equation for the gauge field becomes
\begin{equation}
- \partial_{\nu} F^{\mu\nu} + \frac{\kappa}{2} {\epsilon}^{\mu\nu\rho} F_{\nu\rho} = eJ^{\mu},
\label{fci1}
\end{equation}
where the components of $J^{\mu} = (J^{0}, J^{1}, J^{2})$ are given by
\begin{eqnarray}
J^{0} &=& \rho ~~=~~ {|\psi_{1}|}^2,
\label{current0} \\
e \epsilon_{ij} J^{j} &=& -ie \epsilon_{ij} {\psi}^{\dagger} \sigma_{j} \psi - l\partial_{i} \rho.
\label{current}
\end{eqnarray}
Eq.(\ref{fci1}) can be decomposed into three equations:
\begin{eqnarray}
\partial_{i} E^{i} - \kappa B &=& e\rho,
\label{fci20} \\
\epsilon_{ij} \partial_{0} E^{j} + \partial_{i} B - \kappa E^{i} &=& -e \epsilon_{ij} J^{j},
\label{fci2}
\end{eqnarray}
where $E^{i} = - \partial_{0} A^{i} - \partial_{i} A^{0}$, and
$ B = \nabla \times \vec{A} = \epsilon_{ij} \partial_{i} A^{j} $.
Eq.(\ref{fci20}) is the modified Gauss' law.
In the static limit, Eqs.(\ref{fci20}) and (\ref{fci2}) reduce to
\begin{eqnarray}
\partial_{i} {\partial}^{i} A^{0} - \kappa B &=& e\rho,
\label{fci-static0}  \\
\partial_{i} B + \kappa \partial_{i} A^{0} &=& -e \epsilon_{ij} J^{j}.
\label{fci-static}
\end{eqnarray}

Equations of motion for the nonrelativistic matter field can be written as
\begin{eqnarray}
i {\cal{D}}_{t} \psi_{1} + ({\cal{D}}_{1} - i {\cal{D}}_{2}) \psi_{2}
+ 2g ({\psi}_{1}^{\dagger} \psi_{1} ) \psi_{1} &=& 0,  \label{eq of matter10}\\
({\cal{D}}_{1} + i {\cal{D}}_{2}) \psi_{1} - 2m \psi_{2} &=& 0.
\label{eq of matter1}
\end{eqnarray}
Substituting Eq.(\ref{eq of matter1}) into (\ref{eq of matter10}), we obtain
\begin{equation}
\frac{1}{2m} ({\cal{D}}_{1} - i {\cal{D}}_{2})({\cal{D}}_{1} + i {\cal{D}}_{2}) \psi_{1}
+ i {\cal{D}}_{t} \psi_{1} + 2g ({\psi}_{1}^{\dagger} \psi_{1} ) \psi_{1} = 0,
\label{schr eq1}
\end{equation}
which is the second-order Schr\"odinger equation modified by the introduction of the gauge
coupling and the self-interaction term.
In the static limit, ${\cal{D}}_{t}$ reduces to ${\cal{D}}^s_{t} = ieA^{0} - ilB$.

In order to find the static vortex solutions we choose the spinor field as
\begin{equation}
\psi = \Bigl{(}
\begin{array}{cc}
{\psi}_{1} \cr 0   \nonumber
\end{array} \Bigr{)} ,
~~~~~{\psi}_{2} = 0.
\label{psifieldchoice}
\end{equation}
Then Eq.(\ref{eq of matter1}) reduces to
\begin{equation}
({\cal{D}}_{1} + i {\cal{D}}_{2}) \psi_{1} = 0,
\label{self-dual condition}
\end{equation}
which is the self-dual equation, and Eq.(\ref{eq of matter10}) reduces to an algebraic equation,
\begin{equation}
eA^{0} - lB - 2g\rho = 0.
\label{algebraic eq1}
\end{equation}
Note that, due to this choice of the spinor field, $\psi^{\dagger} \sigma_{i} \psi = 0$ in
Eq.(\ref{current}).

By comparing equations (\ref{fci-static0}) and (\ref{fci-static})
with the definition of the current, Eqs.(\ref{current0}) and (\ref{current}),
we obtain the field equations for the gauge field in the static limit,
\begin{eqnarray}
\partial_{i} {\partial}^{i} A^{0} - \kappa B &=& e\rho ,
\label{fci-current eq10} \\
\partial_{i} B + \kappa \partial_{i} A^{0} &=&  l \partial_{i} \rho .
\label{fci-current eq1}
\end{eqnarray}
Eq.(\ref{fci-current eq1}) can be written as,
\begin{equation}
B + \kappa A^{0} = l \rho.
\label{algebraic eq2}
\end{equation}
If we choose the gauge fixing condition, $A^{0} = 0$, then Eq.(\ref{fci-current eq10})
also reduces to an algebraic equation. For this system to have consistent static solutions,
the algebraic equations (\ref{algebraic eq1}), (\ref{fci-current eq10}) and (\ref{algebraic eq2})
must be identical. Thus the coupling constants $l$, $e$ and $g$ must satisfy the conditions,
\begin{eqnarray}
l &=& - \frac{e}{\kappa},
\label{landg0}    \\
g &=& - \frac{e^2}{2 {\kappa}^2}.
\label{landg}
\end{eqnarray}

The definition of the current, Eq.(\ref{current}), can be written as
\begin{equation}
eJ^{i} = -ie {\psi}^{\dagger} \sigma_{i} \psi + {\epsilon}^{ij0} l \partial_{j} J^{0} .
\label{current3}
\end{equation}
This implies that the current $J^{i}$ includes both the current generated by the matter field,
$-i{\psi}^{\dagger} \sigma_{i} \psi$, and the induced current $G_{ind}^{i}$ from
the magnetic coupling,
\begin{eqnarray}
G_{ind}^{i} &=& \frac{l}{e} {\epsilon}^{ij0} \partial_{j} J^{0}   \nonumber   \\
&=& - \frac{1}{\kappa} {\epsilon}^{ij0} \partial_{j} J^{0}.
\label{induced current}
\end{eqnarray}
There is no induced charge density because we have coupled the anomalous magnetic terms
only through the time component of the covariant derivative.
Due to the choice of the matter field, Eq.(\ref{psifieldchoice}),
the matter field current, $-i{\psi}^{\dagger} \sigma_{i} \psi$, vanishes
and the charge density $\rho = J^{0}$ remains stationary in the static limit.

Barci and Oxmen \cite{barci} have shown that, at every point where we have a charge $Q$ and a magnetic
dipole moment $\mu \hat{z}$ ($\hat{z}$ is a unit vector perpendicular to the plane in
consideration), the magnetic dipole moment density is given by
\begin{equation}
\vec{m}(\vec{x}) = \frac{\mu}{Q} J^{0} \hat{z},
\label{magnetic dipole moment density}
\end{equation}
in the case where the charge density is stationary and current density vanishes.
This dipole moment density induces an electric current (in (2+1)-dimensional notation);
\begin{equation}
G_{ind}^{i} = \frac{\mu}{Q} {\epsilon}^{ij0} \partial_{j} J^{0}.
\label{induced current from BO}
\end{equation}
By comparing this induced electric current with that of our model,
we find
\begin{equation}
\frac{l}{e} = - \frac{1}{\kappa} = \frac{\mu}{Q}.
\label{lkappamu relation}
\end{equation}
We thus find that there exists a magnetic dipole moment, $\mu = - \frac{Q}{\kappa}$,
in this system. This anomalous magnetic coupling can be thought of
the (2+1)-dimensional reduction of the familar Pauli magnetic moment
coupling in (3+1)-dimensions \cite{kogan}.

The solutions of the self-dual equation (\ref{self-dual condition}) is well-known and given by
\begin{equation}
B = \epsilon_{ij} \partial_{i} A^{j} = -\frac{1}{2e} {\nabla}^2 \ln \rho.
\label{solofB}
\end{equation}
Substituting this into the Gauss' law constraint, Eq.(\ref{fci-static0}),
we obtain the Liouville equation for the charge density:
\begin{equation}
{\nabla}^2 \ln \rho = \frac{2e^2}{\kappa} \rho,
\label{inhomogeneousLiouville1}
\end{equation}
where $\kappa$ must be negative for the regularity of the matter density.
The solutions of the Liouville equation is well-known\cite{Jackiw},
and this shows that the nonrelativistic fermion model described by the Lagrangian (\ref{lagrangian1})
supports the static vortex solutions.

We now consider the second possibility of choosing the nonsingular matrix $\beta$ of
Eq.(\ref{twobetachoice}). If we take
\begin{equation}
\beta = \left( \begin{array}{cc} -1 & 0 \\ 0 & -1 \end{array} \right),
\label{betachoice2276}
\end{equation}
$B_{4}$ becomes $B_{4} = \left( \begin{array}{cc} i & 0 \\ 0 & i \end{array} \right) $.
From this choice, we obtain the matrices $A$ and $C$ as,
\begin{eqnarray}
A = \left( \begin{array}{cc} 0 & 0 \\ 0 & i \end{array} \right),~~~~
C = \left( \begin{array}{cc} -2im & 0 \\ 0 & 0 \end{array} \right).
\label{ACequation12}
\end{eqnarray}
Then the first-order Schr\"odinger equation becomes
\begin{equation}
\Bigl{(} \frac{i}{2}  (1-\sigma_3 )\partial_t - \vec{\sigma}\cdot\vec{\partial} - m(1+\sigma_3 ) \Bigr{)} \psi = 0.
\label{nonrelativisticeq222}
\end{equation}

We can also construct the Lagrangian for the choice Eq.(\ref{betachoice2276}) of $\beta$
with self-interaction term $g(\psi^{\dagger} \frac{1}{2} (1-\sigma_3 ) \psi)^2 $:
\begin{eqnarray}
{\cal L} = -\frac{1}{4} F_{\mu\nu} F^{\mu\nu}
+ \frac{\kappa}{4} {\epsilon}^{\mu\nu\rho} A_{\mu} F_{\nu \rho}
&+& {\psi}^{\dagger} \Bigl{(} \frac{i}{2} D_t (1-\sigma_3 ) - \vec{\sigma} \cdot \vec{D}
- m (1+\sigma_3 ) \Bigr{)} \psi   \nonumber \\
&+& g \Bigl{(} {\psi}^{\dagger} \frac{1}{2} (1-\sigma_3) \psi \Bigr{)}^2 .
\label{lagrangian2123}
\end{eqnarray}
In order to find the static vortex solutions of this system, we choose the spinor field as
\begin{equation}
\psi = \Bigl{(}
\begin{array}{cc}
0 \cr {\psi}_{2}   \nonumber
\end{array} \Bigr{)} ,
~~~~~{\psi}_{1} = 0.
\label{psifieldchoice2}
\end{equation}
Then the spinor field equation reduces to an antiself-dual equation,
\begin{equation}
({\cal D}_{1} - i{\cal D}_{2} ) \psi_{2} = 0.
\label{anti self-dual eq}
\end{equation}
The gauge field satisfies the same field equations (\ref{algebraic eq1}), (\ref{fci-current eq10}),
and (\ref{algebraic eq2}) as before, so that the coupling constants $l$, $e$ and $g$ satisfy the same
conditions (\ref{landg0}) and (\ref{landg}). Combining these field equations we finally obtain
the Liouville equation,
\begin{equation}
{\nabla}^2 \ln \rho = -\frac{2e^2}{\kappa} \rho,
\label{inhomogeneousLiouville23221}
\end{equation}
for the charge density $\rho = |{\psi}_{2}|^{2}$, where the minus sign comes from the different
self-dual equation and $\kappa$ must be positive in this case for regular vortex solutions.

We have constructed two possible $(2 + 1)$-dimensional nonrelativistic fermionic theories coupled
to a Maxwell-Chern-Simons gauge field that support static vortex solutions. The static matter field
in these models satisfy self-dual or antiself-dual equations depending on the choice of the matrix $\beta$,
which determines the sign of the mass and energy terms in the first-order Schr\"odinger equations,
which in turn determines the sign of the Chern-Simons coupling constant for regular static solutions.

In ref.\cite{eight} it was also shown that two types of static solutions exist for the relativistic
four-fermion theory coupled to a Maxwell-Chern-Simons field, depending on the sign of the Chern-Simons
coupling constant $\kappa$. This implies that the nonrelativistic fermion field equations
(\ref{nonrelativisticeq111}) and (\ref{nonrelativisticeq222}) are somehow related to the nonrelativistic
limits of the relativistic field equations. To see this we consider the free relativistic field equation,
\begin{equation}
i\gamma^{\mu}{\cal D}_{\mu} \Psi - m\Psi = 0,
\label{relativisticeqmotion11}
\end{equation}
where $\gamma^{0} = \sigma^{3}$, $\gamma^{1} = i\sigma^{1}$, and $\gamma^{2} = i\sigma^{2}$.
To obtain the nonrelativistic limit of this equation, we write
$\Psi = e^{\pm imt} \psi$ and use the fact that in the nonrelativistic limit the rest mass is the
the largest energy \cite{drell}. If we take  $\Psi = e^{-imt} \psi $ and use the fact that
there exists an ambiguity in choosing the sign of the matrices $\sigma_{i}$ in our model,
we can obtain the nonrelativistic equation (\ref{nonrelativisticeq111}). If one takes the other choice,
$\Psi = e^{imt} \psi $, on the other hand, one obtains the field equation (\ref{nonrelativisticeq222}).
Thus one can think of two possibility of the matrix $\beta$ for the nonrelativistic fermion field equation,
as different nonrelativistic
limits of the relativistic Dirac equation (\ref{relativisticeqmotion11}).

It is worthwhile to compare the field equations we have derived
and those of the ref.\cite{duval}. Duval et al.\cite{duval} have
derived their field equations from 4-dimensional 4-component
nonrelativistic Dirac equation by dimensional reduction, while we
have constructed the $(2+1)$-dimensional non-relativistic field
equations directly. By integrating out one of the two component
spinor fields ($\chi$, $\Phi$), they obtained the
$(2+1)$-dimensional nonlinear Schr\"{o}dinger equation, thus
deriving the self-interaction term automatically. We have,
however, introduced the self-interaction term in order to cancel
the effects of the newly introduced Maxwell and magnetic
interaction terms to give consistent static vortex solutions.

\section{Nonrelativistic Scalar Field Model}
It is often the case in condensed matter phenomena that spin degrees of freedom in electron
system are frozen and electrons are effectively described as excitations of scalar fields.
For such systems it is meaningful to consider the nonrelativistic scalar field theories that
possess static vortex solutions, and to use them in studying vortex dynamics.
For this purpose we consider a nonlinear Schr\"{o}dinger field theory coupled to a Maxwell-Chern-Simons
gauge field:
\begin{equation}
{\cal L}= - \frac{1}{4}F_{\mu\nu}F^{\mu\nu}+\frac{\kappa}{4}\epsilon^{\mu\nu\rho}A_{\mu}F_{\nu\rho}
+i \phi^*D_t\phi-\frac{1}{2m}|D_i \phi|^2 + \frac{g}{2}(\phi^*\phi)^2,
\label{blag}
\end{equation}
where
\begin{eqnarray}
D_t &=& \partial_0 +i e A_0 + i \frac{l}{2} \epsilon^{ij}F_{ij}\label{tcov},\\
D_i &=& \partial_i + ieA_i,
\label{cov}
\end{eqnarray}
and $\kappa, g$ and $l$ are the coupling constants.
Note that we have also introduced the anomalous magnetic coupling as in the previous case.

The gauge field equations are
\begin{eqnarray}
&&\partial_i E^i -\kappa B = e\rho, \label{gaus1}\\
&&\epsilon_{ij}\partial_0 E^j + \partial_i B - \kappa E^i = -\epsilon_{ij}(eJ^i + l\epsilon_{jl}\partial_l \rho),
\label{geqn}
\end{eqnarray}
where $E^i = -\partial_0 A^i -\partial_i A^0$, $B=\epsilon_{ij}\partial_i A^j$,
$\rho=|\phi|^2$ and $J^j=-\frac{i}{2m}(\phi^* D^j \phi -(D^j\phi)^*\phi)$.
The right hand side of Eq.(\ref{geqn}) includes the matter field
current $J^i$ and the induced one from the magnetic coupling.
Equations of motion for a nonrelativistic matter field is given by
\begin{equation}
i D_t \phi = -\frac{1}{2m}\vec{D}^2\phi -g |\phi|^2\phi,
\label{schro}
\end{equation}
which is the nonlinear Schr\"{o}dinger equation.

In the static limit, the gauge field equations become
\begin{eqnarray}
&&\kappa B =- e\rho, \label{gaus2} \\
&&\partial_i B =-\epsilon_{ij}(eJ^j + l\epsilon_{jl}\partial_l\rho),
\label{seqn}
\end{eqnarray}
where we have chosen the gauge, $A_0 = 0$.
Using the well known identity,
\begin{equation}
\vec{D}^2 \phi =(D_1 \pm iD_2)(D_1\mp iD_2)\phi \pm eB\phi,
\label{iden}
\end{equation}
the nonlinear Schr\"{o}dinger equation (\ref{schro}) can be written as,
\begin{equation}
i\partial_0 \phi= -\frac{1}{2m}(D_1 \pm iD_2)(D_1\mp iD_2)\phi +(\mp \frac{e}{2m}B-lB-g|\phi|^2)\phi.
\label{2schro}
\end{equation}
In the static limit, the Bogomol'nyi limit is saturated by the condition,
\begin{equation}
\mp \frac{e}{2m}B-lB-g|\phi|^2 =0,
\label{alg}
\end{equation}
and Eq.(\ref{2schro}) reduces to the self-dual equations,
\begin{equation}
(D_1\mp iD_2)\phi =0.
\label{self}
\end{equation}

With $\phi$ field satisfying Eq. (\ref{self}) the current density simplifies to
\begin{equation}
J^j =\pm \frac{1}{2m}\epsilon^{jk}\partial_k\rho.
\label{curr}
\end{equation}
For static self-dual solutions, Eq. (\ref{seqn}) becomes
\begin{equation}
\partial_i(B - (\pm \frac{e}{2m} + l)\rho)=0.
\label{alg1}
\end{equation}

For the three equations (\ref{gaus2}), (\ref{alg}) and (\ref{alg1}) to be consistent the coupling
constants must satisfy the conditions,
\begin{eqnarray}
l &=& - \frac{e}{k} \mp \frac{e}{2m},\\
g &=& - \frac{e^2}{k^2}.
\label{const}
\end{eqnarray}
Following the same procedure as in the fermionic case,
i.e. Eqs.(\ref{current3}) $-$ (\ref{induced current from BO}), we obtain the relation
between $l$ and $\mu$,
\begin{equation}
\frac{l}{e} = - \frac{1}{\kappa} \mp \frac{1}{2m} = \frac{\mu}{Q}.
\end{equation}
This implies that the magnetic dipole moment of the system is given by
$\mu = (-\frac{1}{\kappa} \mp \frac{1}{2m})Q$.
The dipole moment $\mu$ is different from the one in the fermionic case by $\mp \frac{1}{2m}Q$.
This difference comes from the difference in the matter field structure. In the fermion models
matter field currents vanish due to the static self-duality condition. In the
scalar model, however, the matter field current survives after imposing the static self-duality condition.
This survived matter field current induces the additional magnetic dipole moment $\mp \frac{1}{2m}Q$.

The self-dual equations (\ref{self}) can be written as
\begin{equation}
B = \pm \frac{1}{2e} \nabla^2 \ln \rho.
\end{equation}
Substituting this into the Gauss' law constraint (\ref{gaus2}), one obtains the well-known Liouville
equation,
\begin{equation}
\nabla^2\ln\rho=\mp \frac{2e^2}{\kappa}\rho,
\label{liou}
\end{equation}
the regular solutions of which exist for $\kappa>0 (\kappa<0)$ for the upper(lower) sign. This model is
free of the singularity problem of the scalar Maxwell-Chern-Simons theory discussed by N\'{e}meth,
\cite{nem} due to the contribution from the anomalous magnetic coupling.

\section{Conclusion}

We have constructed two $(2+1)$-dimensional nonrelativistic fermion field theory models coupled
to a Maxwell-Chern-Simons gauge field, that possess static vortex solutions.
The regular vortex solutions of these models are possible due to the magnetic moment coupling introduced in
addition to the usual minimal coupling to the gauge field. Each of those models supports the regular
vortex solutions for a particular sign of the Chern-Simons coupling constant, similar to the
case of the relativistic fermion Maxwell-Chern-Simons theory \cite{eight}.
We have in fact shown that the first order Schr\"{o}dinger equations satisfied by the fermion matter
field are the nonrelativistic limits of the relativistic Dirac equation.

We have also cinstructed a nonrelativistic scalar field theory coupled to a
Maxwell-Chern-Simons gauge field, that supports static vortex solutions. The regularity of the static solutions
of this model is guaranteed by the magnetic moment coupling, which enables one to
avoid the singularity problem pointed out by Nem\'{e}th\cite{nem}.

Although these models have the same static vortex solutions as those of the Jackiw-Pi model\cite{Pi}
as solutions of the Liouville equation, the moduli space dynamics of the solutions will be quite different
due to the Maxwell term in the Lagrangian.
The reason is that the Maxwell term is quadratic in the time derivatives of gauge fields, i.e.,
$\dot{A_i}\dot{A_i}$, which gives rise to a
non-trivial contributions to the moduli space metric. It would be interesting to study how such a term modify the
low energy dynamics of magnetic vortices.

\begin{center}
{\bf Acknowledgment}
\end{center}
This work was supported in part by Korea Science and Engineering Foundation under Grants No.
97-07-02-02-01-3, by the Center for Theoretical Physics(SNU), and by the
Basic Science Research Institute Program, Ministry of Education, under Project No. 98-015-D00061.
H.-j. Lee was also supported by the Korea Research Foundation
(1997).

\end{document}